\documentclass[fleqn,10pt]{wlscirep}
\title{Electromagnon excitation in cupric oxide measured by Fabry-P\'{e}rot enhanced terahertz Mueller matrix ellipsometry}
\author[1,*]{Sean Knight}
\author[2]{Dharmalingam Prabhakaran}
\author[3]{Christian Binek}
\author[1,4,5]{Mathias Schubert}

\affil[1]{Department of Electrical and Computer Engineering, University of Nebraska-Lincoln, Lincoln, Nebraska, 68588-0511, USA}
\affil[2]{Department of Physics, Clarendon Laboratory, University of Oxford, Parks Road, Oxford OX1 3PU, United Kingdom}
\affil[3]{Department of Physics and Astronomy, University of Nebraska-Lincoln, Lincoln, Nebraska, 68588-0511, USA}
\affil[4]{Terahertz Materials Analysis Center, Department of Physics, Chemistry and Biology, Link\"{o}ping University, SE-58183 Link\"{o}ping, Sweden}
\affil[5]{Leibniz-Institut f\"{u}r Polymerforschung Dresden e.V., Dresden, 01069, Germany}

\affil[*]{sean.knight@engr.unl.edu}
\begin{abstract}
Here we present the use of Fabry-P\'{e}rot enhanced terahertz (THz) Mueller matrix ellipsometry to measure an electromagnon excitation in monoclinic cupric oxide (CuO). As a magnetically induced ferroelectric multiferroic, CuO exhibits coupling between electric and magnetic order. This gives rise to special quasiparticle excitations at THz frequencies called electromagnons. In order to measure the electromagnons in CuO, we exploit single-crystal CuO as a THz Fabry-P\'{e}rot cavity to resonantly enhance the excitation's signature. This enhancement technique enables the complex index of refraction to be extracted. We observe a peak in the absorption coefficient near 0.705~THz and 215~K, which corresponds to the electromagnon excitation. This absorption peak is observed along only one major polarizability axis in the monoclinic \textbf{a}-\textbf{c} plane. We show the excitation can be represented using the Lorentz oscillator model, and discuss how these Lorentz parameters evolve with temperature. Our findings are in excellent agreement with previous characterizations by THz time-domain spectroscopy (THz-TDS), which demonstrates the validity of this enhancement technique.
\end{abstract}
\begin{document}
\flushbottom
\maketitle
\thispagestyle{empty}

\section*{Introduction}
Multiferroic materials are usually defined as materials which exhibit more than one type of ferroic order, for example ferroelectricity and ferromagnetism\cite{erenstein2006multiferroic,ramesh2007multiferroics,cheong2007multiferroics}. This valuable characteristic allows for the implementation of electrical switching of magnetic order, or magnetic switching of electrical order, and which is of interest for potential device applications. One excellent material candidate is cupric oxide (CuO), where ferroelectricity is induced by antiferromagnetic order, i.e., CuO is an induced-multiferroic material\cite{kimura2008cupric}. This characteristic gives rise to a special quasiparticle excitation called the electromagnon. In contrast to a magnon (a spin wave driven by the \textit{magnetic} field of an electromagnetic wave), an electromagnon is a spin wave driven by the \textit{electric} field of an electromagnetic wave\cite{krivoruchko2012electrically}. Electromagnons could provide a means to advance the field of magnonics, in which spin waves are used for information processing\cite{neusser2009magnonics,kostylev2005spin,rovillain2010electric,khitun2009magnetoelectric}. Previously, electromagnons have been identified at low temperatures ($<$~70 K) in multiferroic rare-earth manganites (RMnO$_{\text{3}}$ and RMn$_{\text{2}}$O$_{\text{5}}$)\cite{krivoruchko2012electrically,pimenov2006possible,pimenov2006coupling,sushkov2007electromagnons,kida2009terahertz}, and TbFeO$_{\text{3}}$\cite{stanislavchuk2017far}. However, in CuO electromagnons are seen at relatively higher temperatures (213~K to 230~K)\cite{jones2014high}. To progress towards room temperature multiferroic devices which utilize electromagnons, it is important to investigate materials such as CuO\cite{scott2013room}.

Electromagnons in CuO have been previously characterized by THz time-domain spectroscopy (THz-TDS)\cite{jones2014high,mosley2017terahertz,jones2014influence,mosley2018tracking}. Reference~\citenum{jones2014high} provides a detailed report of the measurement and analysis of this excitation. In Ref.~\citenum{jones2014high}, THz-TDS is used to measure the optical absorption, $\alpha$, of CuO as a function of temperature (200~K to room temperature) and frequency (0.2~THz to 2~THz). When the electric field of the THz beam is parallel to [101] crystal direction, the authors observed a distinct peak in the change of the absorption coefficient $\Delta\alpha$ near 0.73~THz and 214~K. This absorption peak corresponds to the electromagnon excitation.

In general, THz-TDS provides information about the electric field amplitude and phase after interaction with the sample, and therefore allows one to determine the complex-valued refractive index, $\tilde{n}$. THz Mueller matrix ellipsometry is an alternative approach to access $\tilde{n}$ in the THz spectral range\cite{kuhne2014invited,hofmann2012metal,hofmann2017screening,hofmann2013thz,KuehneIEEETHzMMSE2018}. Ellipsometry is a technique which measures the change in the polarization of light after interaction with a sample\cite{fujiwara2007spectroscopic,azzam1987ellipsometry}. An ellipsometric measurement provides information about the relative amplitude and relative phase shift between \textit{s}- and \textit{p}-polarized light, and therefore also grants access to $\tilde{n}$. Since ellipsometry measures relative changes in amplitude and phase, it has the advantage of not depending on the source intensity. The THz ellipsometer system used in this work is described in Ref.~\citenum{kuhne2014invited}. This THz source generates a monochromatic THz beam, in contrast to white-light THz pulses used in THz-TDS. Employing monochromatic THz sources has the benefit of a more direct measurement that does not require an additional step of Fourier-type transforms.

When the THz wavelength ($\lambda \approx 1$~mm) is comparable to the substrate thickness, and when the coherence length of the THz light source is exceeding the substrate thickness by at least one order of magnitude, spectrally-sharp, resonant Fabry-P\'{e}rot interference features can be present in the spectrum of samples deposited onto THz-transparent substrates. This is due to the interference from multiple reflections off the internal front and back interfaces of the substrate. When measured by THz Mueller matrix ellipsometry, these features can be used sensitively to determine the properties of two-dimensional electron gases (2DEGs), for example\cite{hofmann2012temperature,knight2015cavity,armakavicius2016properties,knight2017situ}. In this work, we exploit bulk single-crystal CuO itself as a THz Fabry-P\'{e}rot cavity to enhance the sensitivity to small changes in $\tilde{n}$ as a function of the substrate temperature. A previous report has also demonstrated the use of THz Mueller matrix ellipsometry to identify an electromagnon in TbMnO$_{\text{3}}$\cite{stanislavchuk2013synchrotron}. However, this was accomplished by measuring a single reflection off a bulk single-crystal, and not by exploiting the Fabry-P\'{e}rot enhancement technique described here. For our experimental parameters, a single reflection of the CuO surface would only offer very limited sensitivity to $\tilde{n}$. Our enhancement technique allows accurate characterization of CuO as a function of frequency and temperature in order to observe its electromagnon excitation. We also discuss the application of the Lorentz oscillator model to fit the excitation, and report how these model parameters evolve with temperature. We compare our results with previous investigations by THz-TDS and find excellent agreement.

\section*{Results and Discussion}
\subsection*{Experimental approach}
An illustration of the measurement approach used here is shown in panels (a) and (b) of Fig.~\ref{delta_MM_with_temp}. A thin wafer of single-crystal (010) CuO is exploited as a THz Fabry-P\'{e}rot cavity to enhance the electromagnon's optical signature. The enhancement is caused by interferences between multiple reflections off the front and backside interfaces, as seen in Fig.~\ref{delta_MM_with_temp}. To measure these interferences as a function of temperature and frequency, we employ THz Mueller matrix ellipsometry which provides information about the change in polarization after reflection off the CuO. The measured Mueller matrix data contain very unique features caused by the Fabry-P\'{e}rot interferences. These features are very sensitive to changes in $\tilde{n}$, which enables the characterization of the electromagnon. 

\subsection*{Optical model approach}
\begin{figure}[t]
\centering
\includegraphics[width=\linewidth]{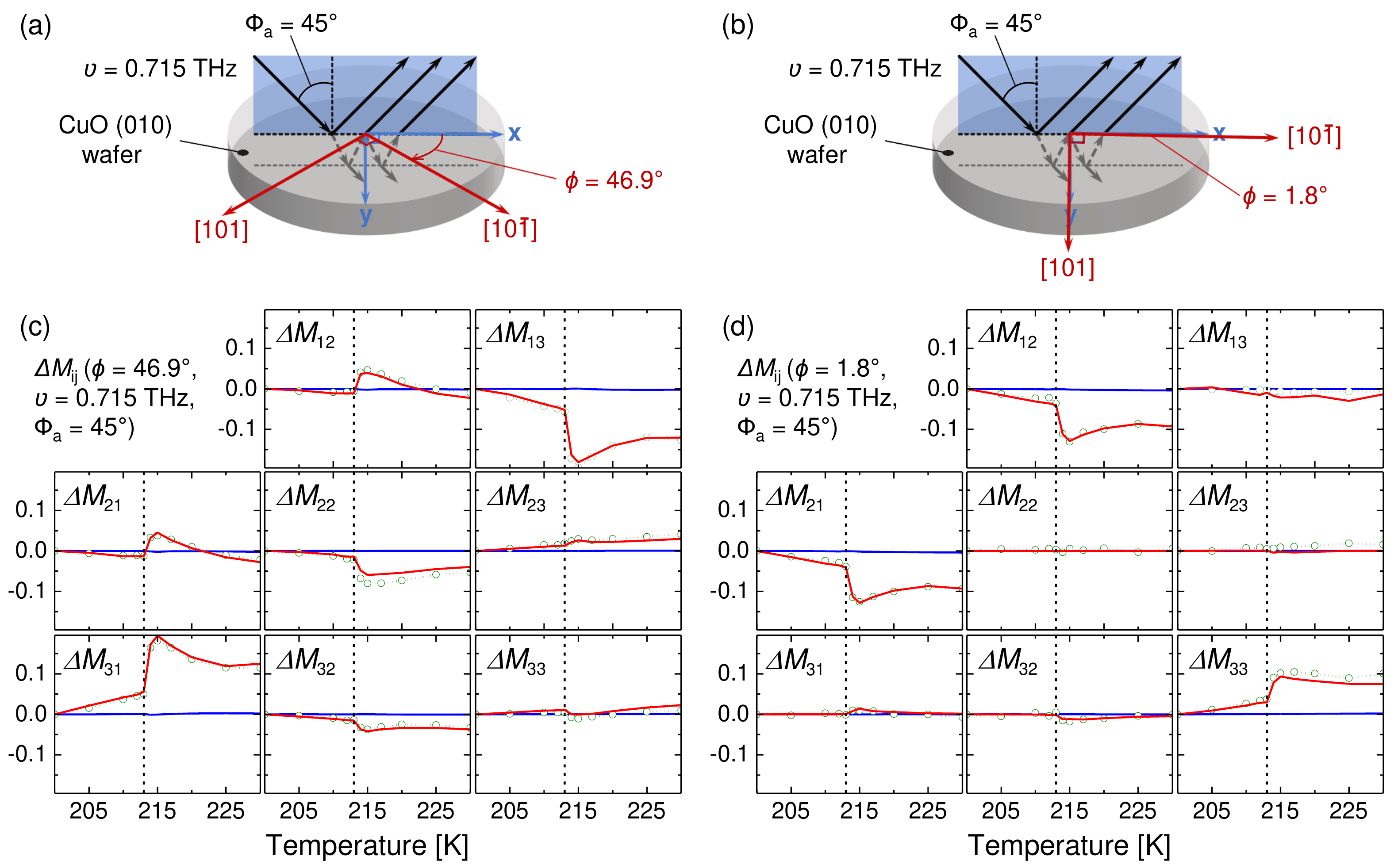}
\caption{Change in the normalized Mueller matrix elements ($\Delta M_{\textrm{ij}} = M_{\textrm{ij}}$($T$)$ - M_{\textrm{ij}}$($T=200~\textrm{K}$)$ $) for single-crystal CuO as a function of temperature at frequency $\nu = 0.715$~THz and at angle of incidence $\Phi_{\textrm{a}} = 45^{\circ}$. Experiment (open green circles with dotted lines) and best-match model calculated data (red solid lines) are for the Fabry-P\'{e}rot enhanced case, in which multiple reflections off the backside of the CuO crystal are included. To demonstrate the extent of the enhancement, simulated non-enhanced data are shown (blue solid lines) where only the first reflection off the CuO surface is considered. Panels (c) and (d) show data for two different azimuth orientations ($\phi = 46.9^{\circ}$ and $\phi = 1.8^{\circ}$, respectively) of the investigated (010) surface cut. Shown in panels (a) and (b) are illustrations of the THz beam's multiple reflections at the front and backside interfaces for each measured azimuth orientation (not to scale). The vertical dashed lines mark the AF1 ($<$ 213~K) to AF2 (213~K to 230~K) phase transition, where CuO becomes an induced-multiferroic in the AF2 phase. This characteristic of the AF2 phase gives rise to the electromagnon excitation. 
}
\label{delta_MM_with_temp}
\end{figure}

The optical model used here consists of a nominally 0.7~mm thick layer of bulk single-crystal CuO with plane parallel interfaces, as shown in Fig.~\ref{delta_MM_with_temp}. The frequency and temperature dependent optical response of CuO is governed by $\tilde{n}(\omega,T)$ which is dependent on the complex permittivity (i.e. dielectric function) $\tilde{\epsilon}(\omega,T)$ and complex permeability $\tilde{\mu}(\omega,T)$ through the equation $\tilde{n}(\omega,T) = \sqrt[]{\tilde{\epsilon}(\omega,T)\tilde{\mu}(\omega,T)}$. As determined in Ref.~\citenum{jones2014high}, we assume $\tilde{\mu}(\omega,T)=1$ for the temperature and frequency range investigated here. The dominant contributions to $\tilde{n}(\omega,T)$ in this range are due to either electromagnons or phonons, both of which behave as electric dipoles, and therefore are represented by $\tilde{\epsilon}(\omega,T)$. Although CuO is a monoclinic crystal, we find the orthorhombic approximation sufficient to fit the measured THz data. For this approximation, we place the three orthogonal major polarizability axes along the [10$\bar{1}$], [101], and [010] crystal directions. This approach was also used by the authors in Ref.~\citenum{jones2014high} to analyze their THz-TDS data. The diagonal Cartesian dielectric tensor used for the orthorhombic approximation is 

%%% Using "gather" should center it, but it does not. Maybe it is not necessary and Sci. Rep. will take care of that?
\begin{equation}
\label{dielectrictensor}
\boldsymbol{\tilde{\epsilon}}= \left(
\begin{array}{ccc}
\tilde{\epsilon}_{\textrm{xx}} & 0 & 0 \\
0 & \tilde{\epsilon}_{\textrm{yy}} & 0 \\
0          & 0 & \tilde{\epsilon}_{\textrm{zz}}
\end{array}\right),
\end{equation}

%%% Seems like Sci. Rep. template is not allowing \boldsymbol{} to work. Need to make eps a tensor (bold it). But \boldsymbol{} works in APL Overleaf template. 
where the tensor elements $\tilde{\epsilon}_{\textrm{xx}}$, $\tilde{\epsilon}_{\textrm{yy}}$, and $\tilde{\epsilon}_{\textrm{zz}}$, are the permittivities along the major polarizability axes [10$\bar{1}$], [101], and [010], respectively. The schematics in Fig.~\ref{delta_MM_with_temp}(a) and \ref{delta_MM_with_temp}(b) include the Cartesian directions \textbf{x} and \textbf{y}, the major polarizability directions [10$\bar{1}$] and [101], and the plane of incidence. The [010] and \textbf{z} directions are omitted for clarity. The direction \textbf{x} is contained within the sample surface plane and oriented along the propagation direction of incident light. The directions \textbf{x}, \textbf{y}, and \textbf{z} are fixed to the THz ellipsometer, while the major polarizability axes (and therefore the CuO crystal) are rotated during the experiment. For the (010) surface cut CuO investigated here, azimuth angle $\phi = 0^{\circ}$ is defined as \textbf{+x} aligned along [10$\bar{1}$]. A positive $\phi$ corresponds to a rotation of the major polarizability axes in the \textbf{a-c} plane in the \textbf{+x} to \textbf{+y} direction. 

\subsection*{Fabry-P\'{e}rot enhanced THz Mueller matrix data}
Shown in Fig.~\ref{delta_MM_with_temp} is the change in the acquired Mueller matrix elements as a function of temperature at a single frequency ($\nu = 0.715$ THz). Data is measured at 205~K, 210~K, 212~K, 213~K, 214~K, 215~K, 217~K, 220~K, 225~K, and 230~K. The Fabry-P\'{e}rot enhanced experimental data (green open circles with dotted lines) and best-match model calculated data (red solid lines) show significant changes as a function of temperature due to the variation in $\boldsymbol{\tilde{\epsilon}}$. The largest change in the data is seen between 213~K and 214~K where the CuO transitions from antiferromagnetic (AF1 phase: $<$ 213~K) to a magnetically induced ferroelectric multiferroic (AF2 phase: 213~K to 230~K). This substantial change in the Mueller matrix is caused by a change in $\boldsymbol{\tilde{\epsilon}}$ due to the appearance of the electromagnon absorption in the AF2 phase. To demonstrate the magnitude of the Fabry-P\'{e}rot enhancement, simulated data for no enhancement effect (blue solid lines) are included in Fig.~\ref{delta_MM_with_temp}. This non-enhanced data is for the case of an infinitely thick CuO crystal, where no reflections off the backside are considered. The $\boldsymbol{\tilde{\epsilon}}(T)$ used to generate the non-enhanced data is determined from the Fabry-P\'{e}rot enhanced data analysis. This analysis will be discussed in detail further below. The non-enhanced data is nearly zero for all temperatures. This shows the Fabry-P\'{e}rot enhancement technique is crucial for obtaining $\boldsymbol{\tilde{\epsilon}}$ in our experiment. 

Two different azimuth orientations of the (010) CuO are measured in our experiments, as illustrated in Fig.~\ref{delta_MM_with_temp}(a) and \ref{delta_MM_with_temp}(b). Figure~\ref{delta_MM_with_temp}(c) shows data for azimuth angle $\phi = 46.9^{\circ}$, and Fig.~\ref{delta_MM_with_temp}(d) shows $\phi = 1.8^{\circ}$. For $\phi = 46.9^{\circ}$, the major polarizability axes in the \textbf{a-c} plane ([10$\bar{1}$] and [101]) have been rotated to near the midpoint between the \textbf{x} and \textbf{y} axes (Fig.~\ref{delta_MM_with_temp}(a)). Since the CuO is anisotropic within the \textbf{a-c} plane, the $\phi = 46.9^{\circ}$ orientation exhibits large \textit{p}-to-\textit{s} and \textit{s}-to-\textit{p} light mode conversion. This mode conversion is quantified by the off-block-diagonal Mueller matrix elements ($M_{\textrm{13}}$, $M_{\textrm{23}}$, $M_{\textrm{31}}$, and $M_{\textrm{32}}$). In contrast, the off-block-diagonal elements for $\phi = 1.8^{\circ}$ are minimal, because the major polarizability axes are near the \textbf{x} and \textbf{y} axes (Fig.~\ref{delta_MM_with_temp}(b)).

\begin{figure}[t]
\centering
\includegraphics[width=\linewidth]{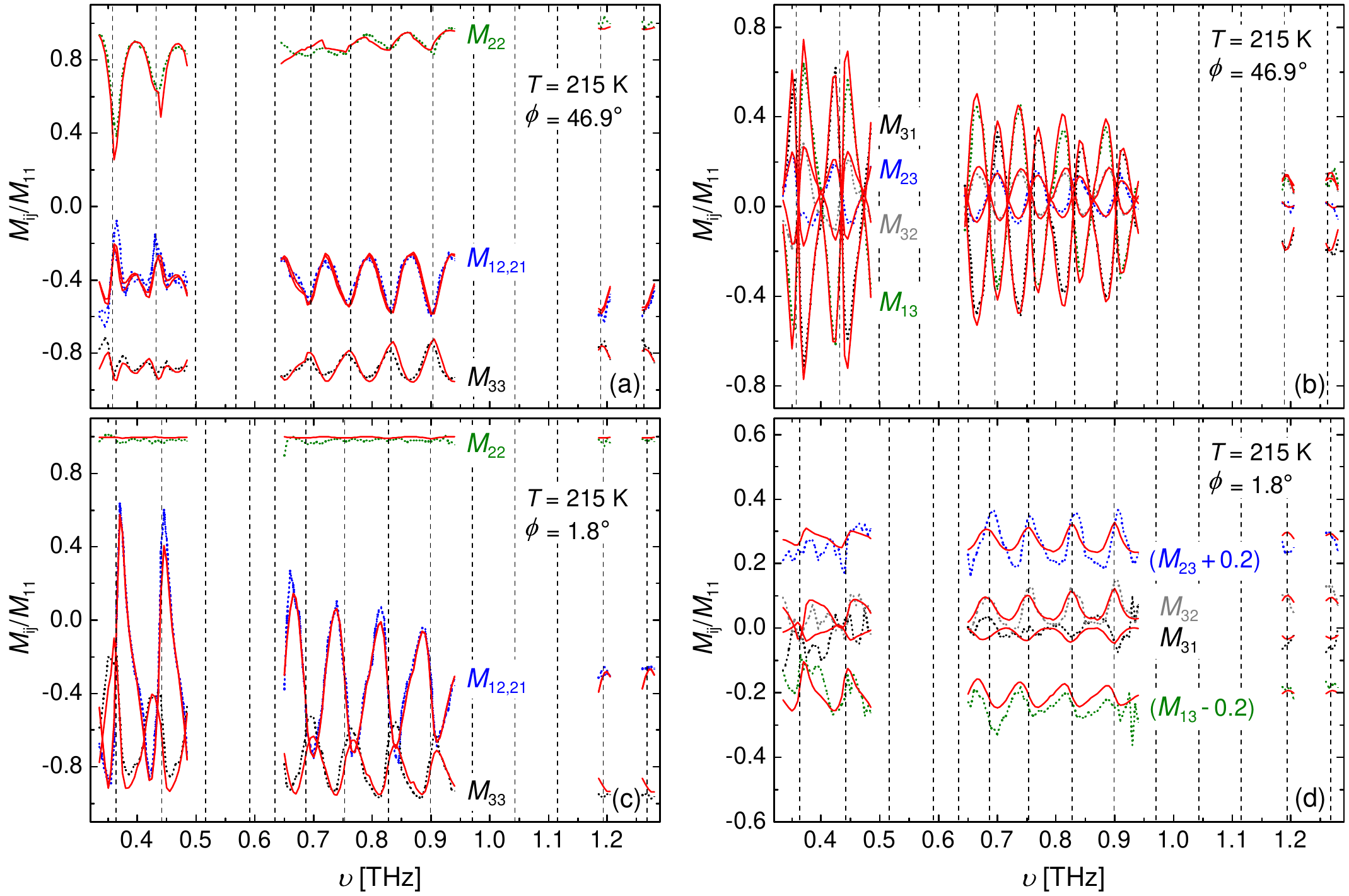}
\caption{Experimental (broken lines) and best-match model calculated (red solid lines) Fabry-P\'{e}rot enhanced Mueller matrix spectra for single-crystal CuO at 215~K and angle of incidence $\Phi_{\textrm{a}} = 45^{\circ}$. Panels (a) and (b) show data from the azimuth orientation $\phi = 46.9^{\circ}$ measurement, and panels (c) and (d) show data for the $\phi = 1.8^{\circ}$ measurement. The acquired on-block-diagonal Mueller matrix elements ($M_{\textrm{12}}$, $M_{\textrm{21}}$, $M_{\textrm{22}}$, and $M_{\textrm{33}}$) are shown in panels (a) and (c), and the off-block-diagonal elements ($M_{\textrm{13}}$, $M_{\textrm{23}}$, $M_{\textrm{31}}$, and $M_{\textrm{32}}$) are shown in panels (b) and (d). Vertical dashed lines indicate the total reflectivity (i.e. $M_{\textrm{11}}$) minima for the respective azimuth orientations.}
\label{On_and_off_diag_spectra}
\end{figure}

Shown in Fig.~\ref{On_and_off_diag_spectra} is the Fabry-P\'{e}rot enhanced Mueller matrix spectra for a single temperature (215~K). Data is measured in increments of 0.005~THz in the available frequency ranges. The minimums in the simulated total reflectivity ($M_{\textrm{11}}$) are shown in Fig.~\ref{On_and_off_diag_spectra} as vertical dotted lines to demonstrate the reflectivity is related to the Mueller matrix. In Fig.~\ref{On_and_off_diag_spectra}, panels (a) and (b) show data for the $\phi = 46.9^{\circ}$ orientation, and panels (c) and (d) show data for $\phi = 1.8^{\circ}$. Here, the Mueller matrix elements are separated into on-block-diagonal (left two panels: (a) and (c)) and off-block-diagonals (right two panels: (b) and (d)). As previously mentioned for Fig.~\ref{delta_MM_with_temp}, due to the orientation of the major polarizability axes in the \textbf{a-c} plane the $\phi = 46.9^{\circ}$ orientation shows sizable off-block-diagonals, whereas $\phi = 1.8^{\circ}$ are minimal. The sharp oscillating features in Fig.~\ref{On_and_off_diag_spectra} are due to Fabry-P\'{e}rot interferences, which are highly sensitive to $\boldsymbol{\tilde{\epsilon}}$, $\phi$, and CuO thickness. The number of oscillations in the spectrum is dependent on $\boldsymbol{\tilde{\epsilon}}$ and CuO thickness. Increasing the CuO thickness causes the number of oscillations to increase, and decreasing the thickness causes the number to decrease. For the CuO sample investigated here, the maximum sensitivity to $\boldsymbol{\tilde{\epsilon}}$ occurs near the reflection minimum of each oscillation. Therefore, a large number of oscillations is desirable to achieve increased sensitivity at as many points in the spectrum as possible. For our experiment, we find a nominal CuO thickness of 0.7~mm is optimal.

\begin{figure}[t]
\centering
\includegraphics[width=\linewidth]{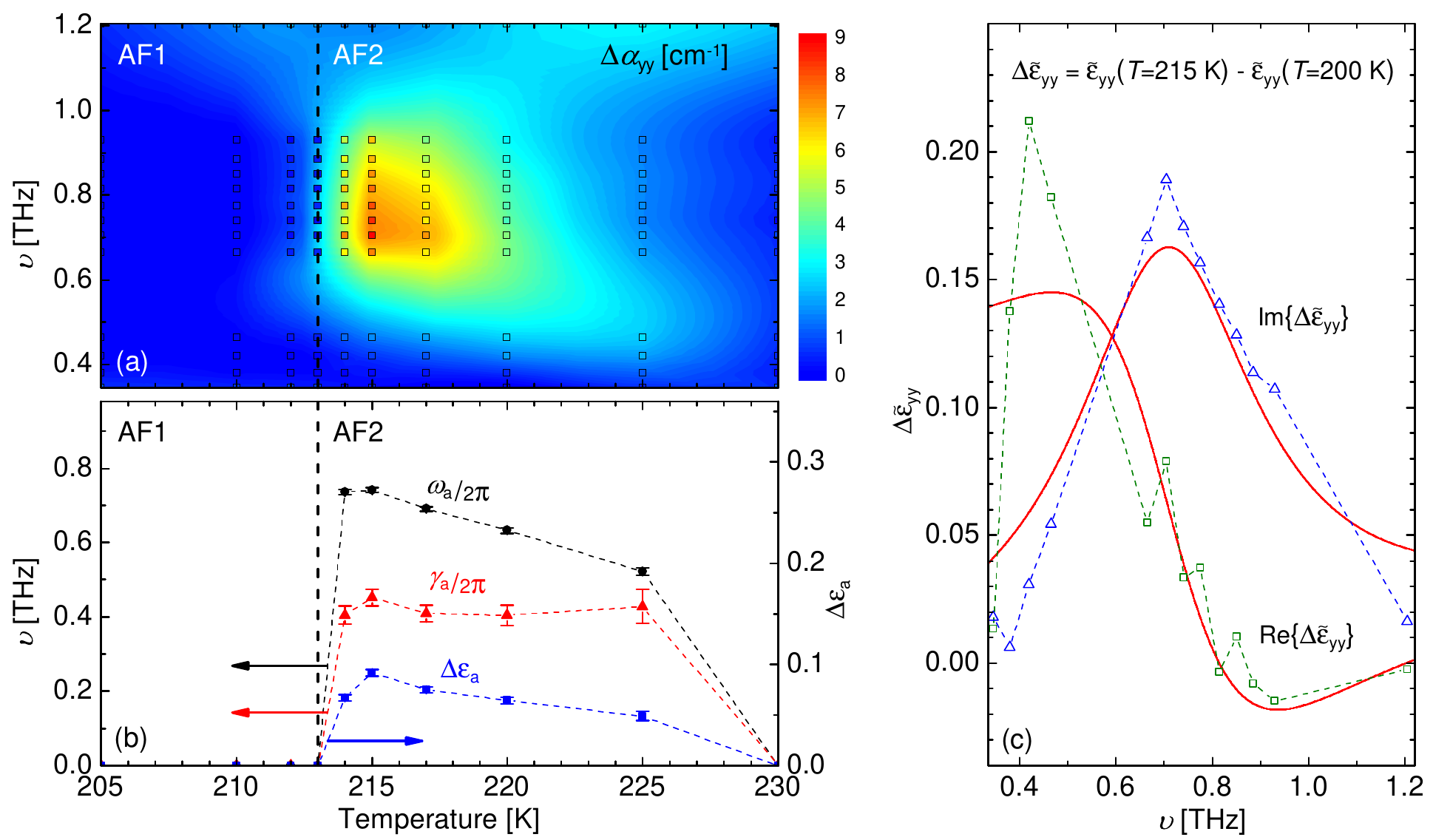}
\caption{Results of the best-match model analysis. Panel (a) shows a false color map of the difference in absorption coefficient ($\Delta\alpha_{yy} = \alpha_{yy}$($T$)$ - \alpha_{yy}$($T=200~\textrm{K}$)$ $) along the [101] major polarizability direction as a function of temperature and frequency. The peak in $\Delta\alpha_{yy}$ near 0.705~THz and 215~K corresponds to the electromagnon excitation. Colored square symbols indicate all individual data points from the piecewise constant fit. On the same color scale is a smoothed contour plot to guide the reader's eye. The dashed line marks the AF1 to AF2 phase transition. Panel (b) shows parameters from the Lorentz oscillator analysis (solid symbols) as a function of temperature. Panel (c) shows an example Lorentz oscillator fit (red solid lines) to the piecewise constant fit values (open symbols) for $\Delta\tilde{\epsilon}_{yy}$ at 215~K. 
}
\label{Results}
\end{figure}

\subsection*{Best-match model analysis results}
Shown in Fig.~\ref{Results} are the results of the best-match model analysis of the Fabry-P\'{e}rot enhanced data. The analysis is performed by employing the optical model approach described previously in this work. To determine $\boldsymbol{\tilde{\epsilon}}$ independently for each point in the $(\omega,T)$ array, the Mueller matrix data ($M_{\textrm{ij}}$) and Mueller matrix difference-data ($\Delta M_{\textrm{ij}} = M_{\textrm{ij}}$($T$)$ - M_{\textrm{ij}}$($T=200~\textrm{K}$)$ $) for all measured temperatures and frequencies are analyzed simultaneously. With respect to frequency, multiple data points are grouped together and assigned the same value for $\boldsymbol{\tilde{\epsilon}}$ in the analysis. The bounds for these sections in units of THz are: 0.360, 0.402, 0.438, 0.600, 0.687, 0.720, 0.755, 0.794, 0.830, 0.866, 0.902, and 1.17. This creates 13 independent piecewise sections for which all frequencies in one section have a constant value. We refer to this analysis as the piecewise constant fit approach. The values chosen for the bounds are the Mueller matrix zero-crossings seen in Fig.~\ref{On_and_off_diag_spectra}(b). With respect to temperature, all 10 increments are assigned independent values of $\boldsymbol{\tilde{\epsilon}}$. This piecewise constant fit approach creates a two-dimensional array of values for $\boldsymbol{\tilde{\epsilon}}$ (13 piecewise sections with respect to frequency $\times$ 10 points with respect to temperature). The analysis reveals no features of interest in $\tilde{\epsilon}_{\textrm{xx}}(\omega,T)$. Due to limited sensitivity in the \textbf{z} direction, we set $\tilde{\epsilon}_{\textrm{zz}}(\omega,T)$ to a constant value in the analysis (see \textbf{Methods} for further details). However, for $\tilde{\epsilon}_{\textrm{yy}}(\omega,T)$, a distinct peak in the absorption coefficient, $\alpha_{\textrm{yy}}=2\frac{\omega}{c}Im\{\sqrt{\tilde{\varepsilon}_{\textrm{yy}}}\}$, is seen, and which corresponds to the electromagnon excitation. Figure~\ref{Results}(a) shows a false color map of the difference in the absorption coefficient $\Delta\alpha_{\textrm{yy}}$ along the [101] direction ($\Delta\alpha_{\textrm{yy}} = \alpha_{\textrm{yy}}$($T$)$ - \alpha_{\textrm{yy}}$($T=200~\textrm{K}$)$ $). The peak seen near 0.705~THz and 215~K corresponds to the electromagnon excitation. A sharp increase in $\Delta\alpha_{\textrm{yy}}$ is observed from 213~K to 214~K due to the sudden appearance of the electromagnon in the AF2 phase. 

Since electromagnons primarily behave as electric dipoles, the Lorentz oscillator has been used to model their optical response\cite{jones2014high,stanislavchuk2013synchrotron}. The electromagnon excitation in CuO has been previously modeled by using the sum of two Lorentz oscillators\cite{jones2014high}

% ?Possibly use \left \right to make "()" sizes change?
\begin{equation}\label{Lorentz}  
\Delta\tilde{\epsilon}_{\textrm{yy}}(\omega,T) = \tilde{\epsilon}_{\textrm{yy}}(\omega,T)-\tilde{\epsilon}_{\textrm{yy}}(\omega,T=200~\textrm{K}) = \frac{\Delta\epsilon_{\textrm{a}}\cdot\omega^2_{\textrm{a}}}{\omega^2_{\textrm{a}}-\omega^2-i\omega\gamma_{\textrm{a}}} + \frac{\Delta\epsilon_{\textrm{b}}\cdot\omega^2_{\textrm{b}}}{\omega^2_{\textrm{\textrm{b}}}-\omega^2-i\omega\gamma_{\textrm{b}}},
\end{equation}

where $\Delta\epsilon_{\textrm{a,b}}$, $\omega_{\textrm{a,b}}$, and $\gamma_{\textrm{a,b}}$ are the amplitude, center frequency, and broadening parameters for each mode, respectively. This model is fit to the \textit{change} in the dielectric function $\Delta\tilde{\epsilon}_{\textrm{yy}}$ relative to 200~K in an attempt to isolate the electromagon and lessen the contributions from phonon modes\cite{jones2014high}. Mode $a$ is the main electromagnon mode, and mode $b$ is a broad low-amplitude shoulder mode to the electromagnon. We fit the mode $a$ parameters in Eqn.~\ref{Lorentz} to the values of $\Delta\tilde{\epsilon}_{\textrm{yy}}(\omega,T)$ from the piecewise constant fit to obtain the Lorentz oscillator parameters as a function of temperature. Due to the limits of our available spectral range, the mode $b$ parameters were fixed to values determined in Ref.~\citenum{jones2014high}. Figure~\ref{Results}(b) shows the results of the Lorentz oscillator model analysis for the main electromagnon mode $a$. The $\Delta\epsilon_{\textrm{a}}$ and $\omega_{\textrm{a}}$ parameters show similar trends of a dramatic increase from 213~K to 214~K followed by a gradual decrease to zero. Note, $\omega_{\textrm{a}}(T)$ does not exactly coincide with maximum for $\Delta\alpha_{\textrm{yy}}(T)$, because $\gamma_{\textrm{a}}$ is comparable to $\omega_{\textrm{a}}$. The $\gamma_{\textrm{a}}$ parameter seems to follow a similar trend until 220~K and 225~K. Since the absorption peak is beginning to move outside the available spectral range at 220~K, it is more difficult to determine $\gamma_{\textrm{a}}$, which is reflected in the larger error bars for the 220~K and 225~K data points. These results are in excellent agreement with the parameters provided in Ref.~\citenum{jones2014high} where the Lorentz model is fit to only $\Delta\alpha_{\textrm{yy}}$, instead both the real and imaginary parts of $\Delta\tilde{\epsilon}_{\textrm{yy}}$ as in this work. Shown in Fig.~\ref{Results}(c) is an example of the Lorentz oscillator best-match model fit to the piecewise constant fit values for $\Delta\tilde{\epsilon}_{\textrm{yy}}(\omega,T=215~\textrm{K})$. We note electromagnons can also contribute to the magnetoelectric tensors (i.e., gryotropic tensors, or cross tensors), which enable dynamic electric influence of magnetic polarization, and dynamic magnetic influence of electric polarization\cite{stanislavchuk2013synchrotron}. For example, Ref.~\citenum{stanislavchuk2013synchrotron} discusses the characterization of an electromagnon in single-crystal TbMnO$_{\text{3}}$ in which a small contribution to one of the magnetoelectric tensors is modeled using a Lorentz oscillator. In general, it is possible to use the Mueller matrix to differentiate contributions in $\boldsymbol{\tilde{\epsilon}}$, $\boldsymbol{\tilde{\mu}}$, and the magnetoelectric tensors\cite{rogers2011mueller,stanislavchuk2013synchrotron}. However for CuO, a more rigorous analysis is needed considering its complex monoclinic nature. 

\section*{Conclusion}
A Fabry-P\'{e}rot enhanced terahertz (THz) Mueller matrix ellipsometry approach was used to determine the electromagnon excitation in monoclinic cupric oxide (CuO). A single-crystal CuO cut with parallel interfaces was exploited as a THz Fabry-P\'{e}rot cavity to resonantly enhance the excitation's signature. This enhancement technique enables the complex index of refraction to be extracted. We observe a peak in the absorption coefficient near 0.705~THz and 215~K, which corresponds to the electromagnon excitation. Our findings are in excellent agreement with previous characterizations by THz time-domain spectroscopy (THz-TDS). We propose the use of the THz enhancement technique to detect small absorption changes in anisotropic crystals caused by subtle excitations such as electromagnons.

\section*{Methods}
\subsection*{Experimental setup and procedure}
The THz ellipsometer sub-system described in Ref.~\citenum{kuhne2014invited} is used to measure bulk single-crystal CuO as a function of temperature and frequency. The THz ellipsometer operates in the polarizer-sample-rotating-analyzer configuration which allows access to the upper-left 3$\times$3 block of the complete 4$\times$4 Mueller matrix. All Mueller matrix data shown here has been normalized to the $M_{11}$ element. The THz source is a backward wave oscillator (BWO) equipped with GaAs Schottky diode frequency multipliers. The detector is a liquid helium cooled bolometer. The magneto-cryostat sub-system is used to cool the sample, but no magnetic field is applied at any point during the experiment. Inside the cryostat, the sample was always measured starting from the lowest temperature (200~K) up to the highest temperature (297~K).

\subsection*{Sample growth}
One disk-like wafer (nominally 8~mm in diameter and 0.7~mm thick) of single-crystal (010) CuO was grown using the optical float zone method\cite{prabhakaran2003single}. Polycrystalline cylindrical feed rods were prepared using high purity (99.995\%) CuO starting chemical and sintered at 900$^{\circ}$~C for three days under oxygen flow. Single-crystal was grown using a four mirror optical floating-zone furnace under 9 bar oxygen pressure\cite{prabhakaran2003single}. The growth was carried out using a sintered feed rod at a growth rate of 3.5~mm/h with feed and seed rods counter-rotating at 30 rpm. One cylindrical disk of (010) single-crystal was cut from the large as-grown crystal for these experiments.

\subsection*{Room temperature THz analysis}
At room temperature outside the cryostat, Fabry-P\'{e}rot enhanced THz Mueller matrix data were taken to determine the unknown sample parameters: CuO wafer thickness, and $\theta$ (rotation of the major polarizability axes about the [10$\bar{1}$] direction). Considering only one angle of incidence is available when measuring through the cryostat ($\Phi_{\textrm{a}} = 45^{\circ}$), these parameters must be obtained outside the cryostat. These measurements were performed at multiple angles of incidence ($\Phi_{\textrm{a}} = 40^{\circ}$, $50^{\circ}$, and $60^{\circ}$), at four azimuth orientations (nominally $\phi = 0^{\circ}$, $45^{\circ}$, $90^{\circ}$, and $135^{\circ}$), and in the frequency range of 0.65~THz to 0.9~THz in increments of 0.005~THz. All the data is analyzed simultaneously to find the CuO thickness is $(0.669\pm 0.003)$~mm, and $\theta = (5.2\pm 0.6)^{\circ}$. Change in the CuO thickness with temperature is set in the optical model according to expansion coefficients reported in Ref.~\citenum{rebello2013multiple}. $\theta$ is fixed in the analysis for all temperatures as it does not depend on $\phi$, temperature, or any other experimental variables. Values of $\phi$ for each azimuth orientation measured through the cryostat are determined by applying the optical model for outside the cryostat. The values for $\phi$ when the sample is mounted in the cryostat are found to be $\phi = (46.9\pm 0.5)^{\circ}$ and $\phi = (1.8\pm 0.6)^{\circ}$.

Analysis of data measured outside the cryostat also allows the room temperature $\boldsymbol{\tilde{\epsilon}}$ to be extracted. At room temperature CuO exhibits minimal dispersion from 0.65~THz to 0.9~THz\cite{jones2014high}, therefore we assume constant values for $\boldsymbol{\tilde{\epsilon}}$ in this range. We find the tensor elements of $\boldsymbol{\tilde{\epsilon}}$ near 0.775~THz are: $\tilde{\epsilon}_{\textrm{xx}} = (10.56\pm 0.09) + i(0.31\pm 0.01)$, $\tilde{\epsilon}_{\textrm{yy}} = (9.64\pm 0.08) + i(0.17\pm 0.01)$, and  $\tilde{\epsilon}_{\textrm{zz}} = (11.94\pm 0.12) + i(0.33\pm 0.09)$. These results are in excellent agreement with values reported in Ref.~\citenum{jones2014high}. This analysis confirms the validity of our orthorhombic approximation described in the \textbf{Optical model approach} section. Due to limited sensitivity in the \textbf{z} direction for data taken through the cryostat, we fix $\tilde{\epsilon}_{\textrm{zz}}$ to $(11.94 + i0.33)$ for all temperature dependent measurements.

%\bibliography{CuO_electromagnons_bib_file}

\section*{Acknowledgements}
This work was supported in part by the National Science Foundation under award DMR 1808715, by Air Force Office of Scientific Research under award FA9550-18-1-0360, and by the Nebraska Materials Research Science and Engineering Center under award DMR 1420645. M.~S. acknowledges the University of Nebraska Foundation and the J.~A.~Woollam~Foundation for financial support. The authors would like to thank James Lloyd-Hughes, Connor Mosley, and Tino Hofmann for the helpful suggestions and conversations.

\section*{Author contributions statement}
S.K., C.B., and M.S., conceived the experiment. S.K. conducted THz measurements, analyzed data, and drafted the manuscript. D.P. supplied the single-crystal CuO sample. C.B. and M.S. provided laboratory collaboration and theoretical insight. All authors aided in analysis and manuscript preparation.

\section*{Additional information}
\textbf{Competing Interests:} The authors declare that they have no competing interests.

\end{document}